\documentclass[fleqn,usenatbib]{mnras}

\usepackage{newtxtext,newtxmath}
\usepackage[T1]{fontenc}

\DeclareRobustCommand{\VAN}[3]{#2}
\let\VANthebibliography\thebibliography
\def\thebibliography{\DeclareRobustCommand{\VAN}[3]{##3}\VANthebibliography}

\usepackage{graphicx}	
\usepackage{amsmath}	
\newcommand{\kms}{\,km\,s$^{-1}$}

\title[Galaxy mergers increase in cold protocluster]{A first measurement of galaxy merger rate increasing in dynamically colder protoclusters at cosmic noon}

\author[S. Liu et al.]{Shuang~Liu,$^{1,2}$
Xian~Zhong~Zheng,$^{3,1,2}$\thanks{Contact e-mail: \href{mailto:xzzheng@pmo.ac.cn}{xzzheng@sjtu.edu.cn }}
Valentino~Gonzalez,$^{4}$
Xiaohu~Yang,$^{5,3}$
Jia-Sheng~Huang,$^{6,7}$
\and
Dong~Dong~Shi,$^{8}$
Haiguang~Xu,$^{5}$ 
Qirong~Yuan,$^{9}$
Yuheng~Zhang, $^{1,2}$
Run~Wen, $^{1,2}$
Man~Qiao, $^{1,2}$ 
\and
Chao~Yang, $^{1,2}$
Zongfei~Lyu$^{1,2}$
\\
$^{1}$Purple Mountain Observatory, Chinese Academy of Sciences, 10 Yuanhua Road, Nanjing 210023,  China\\
$^{2}$School of Astronomy and Space Science, University of Science and Technology of China, Hefei  230026, China\\
$^{3}$Tsung-Dao Lee Institute and Key Laboratory for Particle Physics, Astrophysics and Cosmology, Ministry of Education, \\ \ \,Shanghai Jiao Tong University, Shanghai, 201210, China\\
$^{4}$Departamento de Astronom\'ia, Universidad de Chile, Camino del Observatorio 1515, Las Condes, Santiago 7591245, Chile \\
$^{5}$Department of Astronomy, School of Physics and Astronomy, Shanghai Jiao Tong University, 800 Dongchuan Road, Shanghai 200240, China \\
$^{6}$Chinese Academy of Sciences South America centre for Astronomy, National Astronomical Observatories, \\ \ \,Chinese Academy of Sciences, Beijing 100101, China \\
$^{7}$CAS Key Laboratory of Optical Astronomy, National Astronomical Observatories, Chinese Academy of Sciences, Beijing 100101, China \\
$^{8}$ Center for Fundamental Physics, School of Mechanics and Optoelectronic Physics, Anhui University of Science and Technology, \\ \ \,Huainan, Anhui 232001, China \\
$^{9}$Department of Physics and Institute of Theoretical Physics, Nanjing Normal University, Nanjing 210023, China \\
}
\date{\today}
\pubyear{2024}

\begin{document}
\label{firstpage}
\pagerange{\pageref{firstpage}--\pageref{lastpage}}
\maketitle

\begin{abstract}
The process of galaxy cluster formation likely leaves an imprint on the properties of its individual member galaxies. Understanding this process is essential for uncovering the evolutionary connections between galaxies and cosmic structures. Here we study a sample of ten protoclusters at $z\sim2$--3 in different dynamical states that we estimate based on spectroscopic data of their members. We combine the dynamical information with HST imaging to measure galaxy sizes and pair fractions. Our analysis reveals a clear anti-correlation between the velocity dispersion of the protocluster and its galaxy pair fractions (indicative of merger rates). The velocity dispersion also anti-correlates with the dispersion in size among of the member galaxies. These correlations may be explained by protoclusters in colder dynamical states maintaining a velocity dispersion and galaxy number density that boosts galaxy mergers, which in turn contributes to the structural expansion and compaction of galaxies. Our findings offer constraints for cosmological models regarding the evolution of galaxy morphology across different stages in the assembly of protoclusters.

\end{abstract}

\begin{keywords}
galaxies: clusters: general 
--- galaxies: high-redshift
--- galaxies: evolution
--- galaxies: interactions 
--- galaxies: statistics
\end{keywords}



\section{Introduction}

In the Lambda cold dark matter ($\Lambda$CDM) paradigm, galaxies form in dark matter (DM) haloes grow hierarchically through mergers with other haloes in the large-scale structures of the Universe, known as the cosmic web, consisting of  voids, sheets, filaments, and clusters \citep[e.g.][]{Cole+2000, Peacock2001, Mo2010}.  The growth of these structures is driven by gravity, acting against the cosmic expansion. The eventual collapse of matter into singular haloes is modulated both by the local matter density relative to the cosmic mean (i.e. overdensity), and the surrounding tidal fields \citep{Bond1996, Springel2006, Frenk2012}. While most previous studies have focused on understanding physical processes driving galaxy growth and regulating baryonic matter recycling at (sub)galactic scales within their haloes \citep{Dave2007,Mitra2015,Somerville2015,Christensen2016,Donahue2022,Wright2024}, how the assembly of the cosmic structures impacts galaxy evolution remains largely unexplored, especially during the early formation stages \citep{Ocvirk2008, Madau2014, Overzier2016, Alberts2022,Yajima2022,Rennehan2024,Szpila2024}.

Cosmic noon ($z\sim 2$--4) represents the peak epoch for the formation of cosmic structures and massive galaxies. During this period, protoclusters of galaxies, i.e. the progenitors of massive galaxy clusters observed in the present-day Universe, are typically situated at the intersection nodes of dense, gas-rich filaments within the cosmic web. These filamentary structures critically shape the spatial distribution of accretion onto dark matter haloes and the frequency of galaxy mergers, resulting in a variety of dynamical states within these protoclusters \citep{White2010, Crain2023}.  Understanding the relationship between galaxy properties and the dynamical state of protoclusters sheds light on the early phases of cluster assembly, and the complex interplay of gravitational and baryonic processes that deeply influence their evolution. 

With the current generation of observational facilities, it remains challenging to spectroscopically map galaxies in protoclusters and to link galaxy properties with both their local and global dynamics over scales of $\sim$10--30 comoving Mpc at $z\sim2$--4 \citep{Zheng2021, Shi2021}. Due to their high galaxy densities and relatively low velocity dispersions ($<$500\,\kms), galaxy groups often exhibit frequent interactions, mergers, and efficient gas accretion \citep{Tempel2014}. Within the assembly of protoclusters, it is still unclear to what extent the properties of galaxies are shaped by the `pre-processing' that occurs in the infalling galaxy groups embedded within the surrounding filamentary structures. This pre-processing can enhance the galaxy merger rate, thereby stimulating starbursts and morphological transformations \citep{McIntosh2008, Tran2008, Kocevski2011}. Previous measurements on galaxy merger rates in protoclusters have yielded inconsistent results --- some find the rates to be higher \citep{Lotz2013, Hine2016, Coogan2018, Watson2019, Polletta2021, Liu2023}, while others report rates consistent with typical values found in general fields \citep{Delahaye2017, Monson2021}. This discrepancy may stem from the different dynamical states of the protoclusters under investigation. Consequently, further studies are necessary to examine the relationship between galaxy merger rates and the velocity dispersion of protocluster galaxies, which will help unravel the complex processes related to the evolution of protoclusters.

Moreover, the evolution of galaxy sizes is governed by the angular momentum derived from large-scale tidal torques \citep[e.g.][]{Jiang2019}. Discs may expand as they accrete gas from cold streams \citep{Dekel2009, Pillepich2019, ForsterSchreiber2020}. Conversely, significant structural changes such as major mergers \citep{Kretschmer2020, Tacconi2020}, the incorporation of counter-rotating streams, and disc instabilities \citep{DB2014, Zolotov2015} within protoclusters can induce compaction in galaxies. Studies of galaxies within protoclusters suggested that their sizes can be relatively smaller \citep{Kuchner2017}, larger \citep{Afanasiev2023}, or display no significant difference \citep{Naufal2023} compared to their counterparts in field galaxies. The divergent evolutionary paths of protocluster galaxies with different stellar masses may account for these size variations \citep{Xu2023}. Additionally, such size discrepancies might be attributed to the different dynamical states of overdense regions \citep{Liu2023}. Investigating the size distribution of galaxies in protoclusters with varying dynamical states could shed light on the nature of these discrepancies.

In this study, we collect a pilot sample of protoclusters with spectroscopy for more than 10 member galaxies, and high-resolution, deep imaging from the \textit{Hubble Space Telescope} (HST), which enables us to determine galaxy morphologies and sizes. Our goal is to present an inventory of the pair fraction in evolving protoclusters of a diverse range of dynamics. The compilation of our sample of protoclusters is detailed in Section~\ref{sec2:data}. The computation of their velocity dispersion, merger rate, galaxy size, and our main results about the relationship between pair fraction, galaxy size and velocity dispersion are presented in Section~\ref{sec3:results}. We discuss and summarize our results in Section~\ref{sec4:discussion and summary}.

\section{Sample and Data}\label{sec2:data}

Protoclusters at $z\sim2$--3 serve as key laboratories for understanding the rapid evolution of galaxies and its connection with the large-scale environments. 
However, only a few high-mass protoclusters have extensive coverage of spectroscopical observations and high-resolution imaging data.
In this section, we introduce our methodology for selecting massive protoclusters in the COSMOS field and the collection of additional protoclusters in the literature. 


\subsection{Protoclusters at $z=$2--3 in the COSMOS field}\label{sec2.1}

The COSMOS field, with an area of $\sim$2\,deg$^2$, is one of the survey fields with deepest multiwavelength observations. The latest photometric catalogue, COSMOS2020, by \cite{Weaver2022}, contains roughly one million galaxies with accurate measurements of physical parameters. This field has the widest coverage by the HST WFC3/F160W imaging from the 3D-DASH survey \citep[see][for more details]{Mowla2022}. A number of  followup programs have provided spectroscopic observations for identification of protoclusters at $z\sim2$--3.  

Several protoclusters have been confirmed at $2<z<3$ in the COSMOS field, including 
(1) one at $z=2.2$ with a velocity dispersion of $645\pm 69$\kms \citep{Sobral2013,Sobral2014,Darvish2020}; 
(2) a \textit{Planck} selected protocluster PHz\,G237.01+42.50 at $z=2.16$ \citep{Polletta2021}; 
(3) the most distant X-ray cluster CLJ1001 at $z=2.51$, but filled with massive CO(1-0) emitters \citep{Wang2016,Wang2018}; 
(4) PCL1002 \citep{Casey2015} and COS1000 \citep{Diener2015} in combination with CLJ1001, to form a proto-supercluster,  dubbed as ``Hyperion'' \citep{Cucciati2018}.  

We also collect catalogues of spectroscopic redshifts (spec-$z$) from the literature to increase the number of confirmed member galaxies in the selected protoclusters. Here we list the programs that provide new  spec-$z$ for galaxies at $2<z<3$  in COSMOS: 
\begin{enumerate}
\item[(1)] the VIMOS Ultra Deep Survey \citep[VUDS, ][]{LeFevre2015,Tasca2017}; 
\item[(2)] the COSMOS Ly$\alpha$ Mapping And Tomography Observations (CLAMATO) Survey \citep{Lee2018, Horowitz2022};
\item[(3)] the MOSFIRE Deep Evolution Field (MOSDEF) Survey \citep{Kriek2015}; 
\item[(4)] the ZFIRE survey at $1.5<z<2.5$ \citep{Hung2016,Nanayakkara2016}; 
\item[(5)] the bright Lyman-alpha emitters among Spitzer Matching Survey of the UltraVISTA ultra-deep Stripes (SMUVS) galaxies \citep{Rosani2020};
\item[(6)] the PRIsm MUlti-object Survey \citep[PRIMUS, ][]{Cool2013, Coil2011};
\item[(7)] the Complete Calibration of the Color-Redshift Relation (C3R2) Survey \citep{Stanford2021};
\item[(8)] the zCOSMOS survey \citep{Lilly2007, Lilly2009};
\item[(9)] the Deep Im aging Multi-Object Spectrograph (DEIMOS) 10K Spectroscopic Survey catalogue \citep{Hasinger2018};
\item[(10)] the Fiber Multi-object Spectrograph (FMOS)--COSMOS survey \citep{Kashino2013, Zahid2014, Silverman2015}.
\end{enumerate}

From the above programs, we collect 1550 galaxies with spec-$z$ at $2<z<3$. 
We point out that the observed targets in these programs span different redshift ranges and sky regions, and thus the collected galaxies are not homogeneously distributed over the COSMOS field. Galaxies with stellar mass limit
$\log(M_{\rm limit}/{\rm M}_\odot) = 9.14-9.46$ at $z=2-3$ corresponds to a 70\,per\,cent completeness threshold \citep{Weaver2022}. 
We note that this stellar mass cut is used to calculate the completeness from the galaxy stellar mass function \citep[GSFM, ][]{Weaver2023}, assuming that the GSFM applies equally to galaxies in both general fields and overdense structures. 
This stellar mass threshold is then applied to calculate galaxy number density (per arcmin$^2$) across the COSMOS field to be of the general field, by determining the number of galaxies above it within a specified redshift slice, based on the GSFM.
We conservatively set $\log(M_{\rm limit}/M_\odot) = 9.50$ for the redshift range $z=2-3$ in the following analyses.

Here, we describe the method used to select overdense regions in comparison to the general field. 
We first calculate the field number density within each photometric redshift interval of $\Delta z=0.04$ for galaxies with stellar mass $\log(M_*/M_\odot) \geq 9.5$ in the COSMOS2020 catalogue.
Next, for each redshift interval and within a spatial region of $5\arcmin \times 5\arcmin$ ($\sim$2.5$\times$2.5\,Mpc$^2$ at $z=2.5$), we assess the number density in comparison to the field number density to determine the overdensity factor. 
We then select regions where the overdensity factor exceeds 3 and the number of spectroscopic redshifts is greater than 10, identifying four such regions and designating them as protocluster cores.
For each of the four protocluster cores, we identify the densest center and the peak redshift ($z_{\rm peak}$), based on the distribution of spectroscopically confirmed galaxies. 
The protocluster cores in the COSMOS field, named as CC-2.097, CC-2.239, CC-2.475, and CC-2.503, correspond to peak redshifts of $z_{\rm peak}=2.097$, $z_{\rm peak}=2.239$, $z_{\rm peak}=2.475$ and $z_{\rm peak}=2.503$, respectively. 
The number densities of protocluster cores within a redshift range of $z_{\rm peak} \pm 0.02$ are 1.760, 0.960, 1.480, and 0.960 per arcmin$^ 2$, in comparison to field number densities of 0.207, 0.209, 0.210, and 0.210 per arcmin$^2$ within the same redshift slices. 
The locations of the protocluster cores and their $z_{\rm peak}$ values are listed in Table.~\ref{tab1}.


\begin{table*}
\centering
  \caption{The key parameters of a pilot sample of protoclusters at $z=2$--3,  including  the peak redshift of protocluster members ($z_{\rm peak}$), the number of spectroscopically confirmed member galaxies ($N_{\rm spec}$), the number of galaxies with $\log (M_\ast/{\rm M}_\odot) \geq 10.3$ ($N_{\rm m}$), the velocity dispersion of protoclusters given in units of \kms ($\sigma_{\rm los}$), the merger rate ($\Re (\rm Gyr^{-1})$), and the size scatter ($\sigma_{\rm size}$) of protocluster members. }\label{tab1}
  \begin{tabular}{lccccccccl}
  \hline\hline
  {\rm Name} &  R.A. (J2000) &  Decl. (J2000)  &  $z_{\rm peak}$ &  $N_{\rm spec}$ & $N_{\rm m}$  & $\sigma_{\rm los}$   & $\Re ({\rm Gyr^{-1}})$  & $\sigma_{\rm size}$   &  References$^{a}$  \\
 \hline
 CC-2.097  & 10:00:24 &  $+$02:15:18 & 2.097 &  51  &   18  & $672\pm135$ &  $0.10\pm0.07$ &  $0.22\pm0.06$   &  Kriek+15; Nanayakkara+16 \\
 &     &    &     &     &   &    &    &    &   
 Tasca+17; Hasinger+18;  \\
  &     &    &     &     &   &    &    &    &   
 Horowitz+22; Lilly+23 \\
 CC-2.239  & 10:00:50 & $+$02:00:20 & 2.239  &  29   &  9  & $549\pm197$ & $0.00\pm0.02$  & $0.19\pm0.07$    &    Darvish+20  \\
 CC-2.475  & 10:00:24 & $+$02:24:36 &  2.475 &  37  &   12  & $803\pm226$ & $0.00\pm0.09$ & $0.12\pm0.05$   &  Cool+13,Casey+15,  \\
  &     &    &     &     &   &    &    &    &   
Silverman+15; Tasca+17 \\
  &     &    &     &     &   &    &    &    &   
Horowitz+22; Lilly+23 \\
 CC-2.503  & 10:00:55 & $+$02:20:42 &  2.503 &  20   &  16  & $547\pm136$ & $0.43\pm0.11$  &  $0.22\pm0.06$  &  Wang+16; Wang+18  \\
   &    &  &  &    &    &  &   &      &  Horowitz+22   \\
 SSA22-3.067  & 22:17:28 & $+$00:12:02 & 3.067  & 91  & 13 & $342\pm77$ &  $0.15\pm0.04$  &  $0.25\pm0.09$   &   Mawatari+23  \\
 SSA22-3.093  & 22:17:35 & $+$00:17:22 & 3.093  & 235  &  26  & $505\pm62$ & $0.32\pm0.08$  &  $0.25\pm0.06$  &   Mawatari+23\\
 PKS1138-2.160  & 11:40:46 &  $-$26:28:57 & 2.160  & 50 & 24  & $720\pm250$  & $0.40\pm0.06$   &  $0.27\pm0.06$   &    Shimakawa+18\\
   &  &   &   &    &   &   &    &     &    P\'{e}rez-Mart\'{i}nez+23\\
 BOSS1244-2.230  & 12:43:36 &  $+$35:55:12 &  2.230 & 12 & 8 & $565\pm291$  & $0.52\pm0.17$   &   $0.22\pm0.08$  &    Zheng+21; Shi+21; Liu+23 \\
 BOSS1244-2.246  & 12:43:31 & $+$35:53:60  &  2.246 & 29  & 10  & $224\pm63$  &  $0.33\pm0.14$   &  $0.20\pm0.06$   &    Zheng+21; Shi+21; Liu+23    \\ 
 BOSS1542-2.241  & 15:42:52 & $+$38:49:59  &  2.241 & 31  & 13  & $259\pm41$  &  $1.01\pm0.12$     &  $0.29\pm0.08$   &    Zheng+21; Shi+21; Liu+23 \\ 
 \hline
 \end{tabular}
 \begin{flushleft}
    Note. $^{a}$This column provides the abbreviated references. The full references are as follows: \cite{Cool2013, Casey2015, Kriek2015, Silverman2015, Nanayakkara2016, Wang2016, Tasca2017, Hasinger2018, Shimakawa2018, Wang2018, Darvish2020, Shi2021, Zheng2021, Horowitz2022,  Liu2023, Mawatari2023, PerezMartinez2023}. 
\end{flushleft}
\end{table*}

\subsection{SSA22 at $z=3.09$}

The protocluster SSA22 at $z=3.09$ was discovered as an overdensity of Lyman-break galaxies \citep{Steidel1998}, and spectroscopically confirmed to be composed of extended filamentary structures using Ly$\alpha$ emitting galaxies \citep{Hayashino2004, Matsuda2005}. It hosts a giant Ly$\alpha$ emitting nebulae \citep{Matsuda2005},  and contains dusty star-forming galaxies \citep{Umehata2015,Kato2016}, and Active Galactic Nuclei \citep[AGNs,][]{Monson2021,Monson2023}.  The SSA22 H\,{\small  I} Tomography survey \citep[SSA22-HIT, ][]{Mawatari2023} with the DEIMOS instrument on Keck provides the high-$z$ H\,{\small  I} gas tomography through Ly$\alpha$ absorption imprinted in the spectra of background galaxies.  We use a compiled catalogue of 730 galaxies at $z>2$ from \cite{Mawatari2023} to estimate the dynamical states of the substructures of SSA22.  We identify two substructures: one at $3.04<z<3.08$ and the other at $3.08<z<3.12$, named as SSA22-3.067 and SSA22-3.093, respectively. 

In the SSA22 field, there are 16 pointings of HST/WFC3 F160W imaging observations  in the archive (HST proposals with IDs: 13844, 11735, 11636 and 14747).  We use the data to measure morphological parameters for galaxies in SSA22. In Appendix~\ref{secA}, we describe the method used to measure the physical properties, specifically stellar mass of these galaxies.

\subsection{PKS1138 at $z=2.16$}

PKS1138 at $z=2.16$ is a well-studied protocluster, known as ``Spiderweb''. It was found as an overdensity of a variety of galaxy populations, including Ly$\alpha$ emitters \citep[LAEs, ][]{Kurk2000,Venemans2007}, H$\alpha$ emitters \citep[HAEs, ][]{Koyama2013,Shimakawa2018}, CO(1--0) emitters \citep[e.g.][]{Jin2021,Chen2024}, dust-obscured galaxies \citep[e.g. ][]{Dannerbauer2014,Zeballos2018}, X-ray sources \citep[e.g.][]{Tozzi2022a}, and red sequence objects \citep[e.g.][]{Tanaka2010,Doherty2010,Tanaka2013}.
Some of them have been spectroscopically confirmed \citep[e.g.][]{PerezMartinez2023}. Hot gas has been detected in Spiderweb through the thermal Sunyaev-Zeldovich (tSZ) effect, suggestive of being a dynamically active system with $M_{500} = 3.46 \times 10^{13}\,M_{\odot}$ in a `maturing' process \citep{Tozzi2022b,DiMascolo2023}. We make use of the publicly-available spec-$z$ catalogues from \cite{Shimakawa2018} and \cite{PerezMartinez2023} and pinpoint a sample of 50 HAEs associated with the protocluster (PKS1138-2.160). 
We do not find HST/WFC3 F160W imaging observations in PKS1138-2.160, and instead use the F814W imaging data for structural measurements in the following analyses.
We note that the member galaxies in PKS1138-2.160 were mostly selected in the rest-frame UV, compared with the selection based on the rest-frame optical for other protoclusters.  We will discuss the potential effects of this selection on our results  in Section~\ref{sec3:results}.

\subsection{BOSS1244 and BOSS1542 at $z=2.24$}

BOSS1244 and BOSS1542 are two massive protoclusters at $z=2.24$ traced by the groups of coherently strong Ly$\alpha$ absorption imprinted on the spectra of a number of background quasars, and confirmed as the extreme overdensities of HAEs through narrow-band imaging over an area of $20\arcmin \times 20\arcmin$ each \citep{Zheng2021} and followup near-infrared spectroscopy \citep{Shi2021}. BOSS1244 consists of three substructures and BOSS1542 is the first giant filamentary structure  discovered at $z>2$.  An enrichment of extreme starbursts detected at $850\,\micron$ by JCMT/SCUBA2  was reported in both BOSS1244 and BOSS1542 and mostly located in the outskirts of the dense regions of HAEs \citep{Zhang2022}. Interestingly, a pair of massive quiescent galaxies were identified, likely to form a brightest cluster galaxy  in the dense core of BOSS1244 \citep{Shi2024}.  In the two protoclusters, member galaxies are found to show a pair fraction of $22\pm5$ ($33\pm6$)\,per\,cent in BOSS1244 (BOSS1542), being 1.8 (2.8) times higher than the general fields. The cold dynamical states ($<$400\kms) together with the high density are suggested to be responsible for the boosted galaxy merger rates as well as larger scatter in size and S\'ersic index for member galaxies in the two  protoclusters \citep{Liu2023}.  
It is worth noting that in the histogram of spec-$z$ for BOSS1244, two distinct overdensities are identified at $z=2.230$ and $z=2.246$ \citep{Shi2021}.  
We thus subdivide BOSS1244 into two separate structures: BOSS1244-2.230 and BOSS1244-2.246. 
In contrast, BOSS1542 exhibits a single peak in redshift at $z=2.241$, and we refer to this structure as BOSS1542-2.241.

\section{Analysis and Results}\label{sec3:results}

\subsection{Density maps}

\begin{figure*}
\includegraphics[width=18cm]{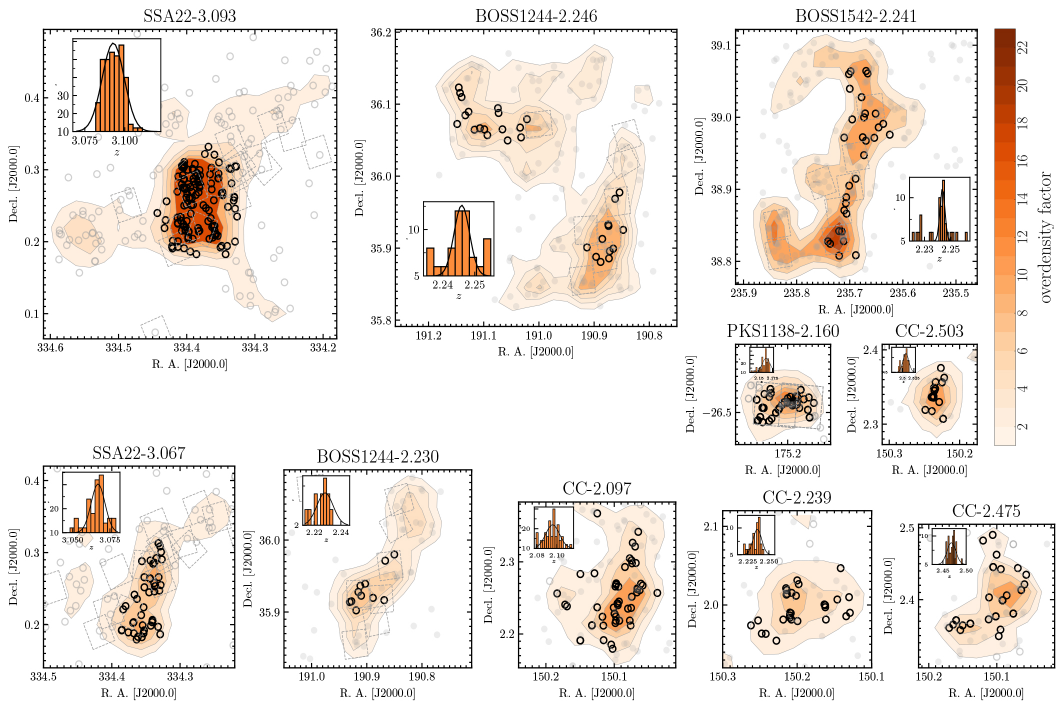}
\caption{Density maps smoothed with a Gaussian kernel with $\sigma = 1$\,arcmin ($\sim$ 1.7\,cMpc at $z=2.5$) of protoclusters at $z=$2--3. The gray dashed lines outline the footprints for HST F160W observation in protoclusters SSA22-3.067, SSA22-3.093, BOSS1244-2.230, BOSS1244-2.246, and BOSS1542-2.241, while for F814W in PKS1138-2.160. 
Contour levels represent [2, 4, 6, 8, 10] times the number density of the general field as described in Section~\ref{sec2.1} for CC-2.097, CC-2.239, CC-2.475, and CC-2.503.
The contour levels are [2, 4, 6, 8, 10] times the average number density of LAEs \protect\citep[0.204 per arcmin$^2$, ][]{Yamada2012} for SSA22-3.093 and SSA22-3.067.
For PKS1138-2.160, the contour levels correspond to [2, 4, 6, 8, 10] times the average number density of HAEs (0.453 per arcmin$^2$) in GOODS-N and UDS-CANDELS fields \protect\citep{Tadaki2011, Tadaki2013}. The contour levels refer to [4, 8, 12, 16, 20] times the number density of HAEs in the general field \citep[0.071 per arcmin$^2$, ][]{An2014} for BOSS1542-2.241, while refer to [2, 4, 6, 8, 10] for BOSS1244-2.230 and BOSS1244-2.246.
The sizes of all panels are scaled to the same standard in terms of their sky coverage. 
Spectroscopically confirmed protocluster members are denoted by open circles, while filled circles represent to the protocluster members without spectroscopic observations. The histogram in each panel shows the distribution of spec-$z$ for member galaxies within protocluster core (denoted by black open circles), and each spec-$z$ bin corresponds to 200\kms at given redshift.}
\label{fig1}
\end{figure*}

Using the collected samples of spectroscopically-confirmed member galaxies, we create density maps for our sample of ten protoclusters.  
The number density of galaxies can be seen as a tracer of the mass density of a protocluster.  
Ideally we would produce the number density maps in the same way for all ten protoclusters, however, the spectroscopic redshift catalogues in SSA22-3.093 and SSA22-3.067 contain more UV-selected member galaxies, while the selection of other protoclusters is closer to a mass selection. 
To determine the density map of a protocluster, it is essential to establish the corresponding number density in the field. The number densities of field galaxies in the COSMOS field are 0.207, 0.209, 0.210, and 0.210 per arcmin$^2$ for CC-2.097, CC-0.239, CC-2.475, and CC-2.503 (as detailed in Section~\ref{sec2.1}).
In Fig.~\ref{fig1}, we show the density map smoothed with a Gaussian kernel of $\sigma = 1$\,arcmin. Protocluster members are further included as those within the contour level that corresponds to twice the field number density for each protocluster in the COSMOS field.
Since the majority ($>60$\,per\,cent) of SSA22 protocluster members are LAEs, we utilize the number density of LAEs in the general fields, 0.204\,arcmin$^{-2}$ \citep{Yamada2012}, to construct relative density maps.
The field number density for PKS1138 is derived from the average number density of HAEs (0.453\,arcmin$^{-2}$) detected in two blank fields: GOODS-N \citep{Tadaki2011} and UDS-CANDELS \citep{Tadaki2013}.
Also, we show the density maps of protocluster members smoothed with a Gaussian kernel of $\sigma = 1$\,arcmin for SSA22-3.093, SSA22-3.067, and PKS1138-2.160 in Fig.~\ref{fig1}. 
The filled color within contours represents the overdensity factors of protoclusters relative to the general field, depicted by different colors in the colormap in Fig.~\ref{fig1}.

We note that the density maps of BOSS1244 and BOSS1542 are determined using the narrow-band selected HAEs after statistically removing background and foreground line emitters \citep[see][for more details]{Zheng2021}.  
The spectroscopic followups revealed that the South-West density region of BOSS1244 consists of two substructures along the line of sight: one peak is at $z\sim2.230$, having 14 galaxies within $z=2.213-2.234$, while the other peaks at $z\sim2.246$ with 9 galaxies in the range $z=2.235-2.255$ \citep{Shi2021}.
For the HAEs without spec-$z$ within the South-West density region, we statistically divide them into the two substructures following a ratio of 14:9 from the distribution of spec-$z$ splited at $z=2.235$.
Following \cite{Zheng2021}, we create density maps smoothed with $\sigma = 1$\,arcmin for BOSS1244-2.230 and BOSS1244-2.246. The reference level for HAEs at $z=2.24$ in the general fields is 0.071\,per\,arcmin$^2$.  
BOSS1542 contains an extended filamentary structure spectroscopically confirmed as a single component in redshift space. Similarly, we adopt the density maps from \cite{Zheng2021}.

Fig.~\ref{fig1} shows the density maps for the ten protoclusters. These protoclusters are shown in the same scale in terms of their sky coverage. We can see from Fig.~\ref{fig1} that our sample protoclusters exhibit a variety of structures, indicative of different dynamical states. The elongated and extended density maps often represent filamentary structures along which material is rapidly accreted into the density core, whereas the density peak regions usually appear as the protocluster core to be virialized first \citep{Casey2016,Ito2019,Rennehan2020,Rennehan2024}. 

\subsection{Velocity Dispersion}

The line-of-sight velocity dispersion ($\sigma_{\rm los}$) is an indicator of total mass for structures, and thus is regarded as a measure of the dynamical state of protoclusters.  We estimate $\sigma_{\rm los}$ using the formula $\sigma_{\rm los} = c\sigma_{z}/(1+z)$, where $c$ represents the speed of light, and $\sigma_z$ denotes the standard deviation of spectroscopic redshifts within each protocluster. 
We estimate $\sigma_z$ by fitting a Gaussian profile to the spectroscopic redshift distribution of protocluster member galaxies. The  redshift histograms and the corresponding best-fitting Gaussian functions are shown in the inner panel of each density map panel  in Fig.~\ref{fig1}. We have also calculated $\sigma_{\rm los}$ using the biweight method for non-Gaussian underlying distributions \citep{Beers1990}, and found that the two methods give consistent results within the uncertainties. 
We note that, due to potential anisotropy in protocluster evolution, velocity dispersion along the line of sight alone may be insufficient to fully capture the dynamical state of the protocluster. A larger sample of high-redshift protoclusters in the future will be necessary to statistically mitigate the impact of this effect.

We present the measured $\sigma_{\rm los}$, the peak redshift ($z_{\rm peak}$), and the number of spectroscopically confirmed protocluster members ($N_{\rm spec}$) for each protocluster in Table~\ref{tab1}. The velocity dispersion $\sigma_{\rm los}$ spans from $\sim$ 200\kms to $\sim$ 800\kms,  indicating a wide range of dynamical states for our sample of protoclusters. Our measurements of $\sigma_{\rm los}$ for SSA22 are consistent with those reported in \cite{Topping2016}, and the high $\sigma_{\rm los}$ for PKS1138-2.160 is supported by the presence of hot intracluster medium indirectly detected through the tSZ effect \citep{DiMascolo2023}.

\subsection{Pair Fraction}\label{sec3.3:pair fraction}

\begin{figure}
  \includegraphics[width=\columnwidth]{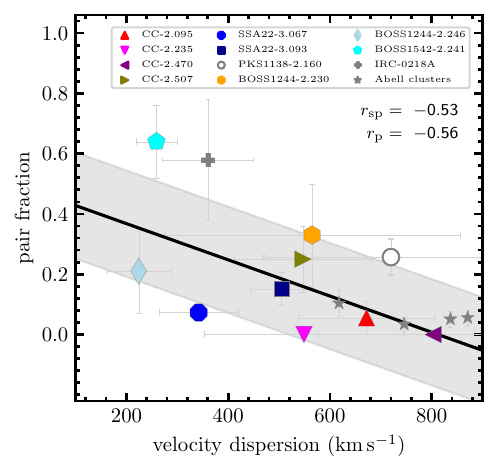}
\caption{Galaxy pair fraction versus velocity dispersion for a pilot sample of ten protoclusters at $z=2$--3. Different protoclusters are represented by various symbols and colors, as illustrated in the legend of this diagram. The parameters of $r_{\rm sp}$ and $r_{\rm p}$ represent to the Spearman and Pearson correlation coefficient between pair fraction and velocity dispersion, excluding PKS1138-2.160 (gray open circle) due to its rest-frame UV images.}
\label{fig2}
\end{figure}

Galaxy major mergers are usually traced by either disrupted morphologies or close pairs. The visual investigation for morphological disturbance is often biased by the intrinsic irregularity of galaxy morphologies and image depth. 
We restrict our sample to $\log(M_\ast/{\rm M}_\odot)\geq10.3$ (we detail the galaxy counts, $N_{\rm m}$ in Table~\ref{tab1})  because massive galaxies with larger haloes tend to have large neighbors that are less influenced by image depth. The following criteria are adopted to count two galaxies as a close pair: (1) a projected distance of 5--30\,kpc; (2) a stellar mass ratio of $\mu>0.25$ if stellar masses are estimated for both of the galaxies, otherwise a flux ratio of $\mu_{\rm flux}>0.25$ is used; (3) the relative line-of-sight velocity difference $\delta_{v}<500$\kms ($\delta_{z}=\delta_{v}\times(1+z)/c$) for spectroscopically confirmed galaxies.  
The selection of $\delta_v=500$\kms ($\delta_z=0.005$ at $z=2$) is conservatively based on the low spectral resolution ($R=200$) in the VUDS survey. This criterion is necessary because, even if the velocity dispersion of a protocluster, such as SSA22-3.067, is less than 500\kms, the spectroscopic redshift difference among its member galaxies can still reach up to 0.03 (equivalent to 3000\kms).
The pair fraction is defined as $f_{\rm pair}=N_{\rm pair}/N_{\rm total}$, where $N_{\rm pair}$ is the number of galaxies with close neighbors and $N_{\rm total}$ is the total number of massive protocluster members. 
For protocluster members without spec-$z$ companions, we lack redshift information for potential companion galaxies, which could lead to contamination from foreground or background sources.
We first search for companion galaxies around each protocluster galaxy using HST images and identify galaxy pairs satisfying the first two criteria (defined as projected pair fraction $f_{\rm proj\ pair}$). 
We then estimate the probability of false pairs ($f_{\rm false\ pair}$) for CC-2.097, CC-2.239, CC-2.475, and CC-2.503 using photometric redshift in COSMOS2020 catalogue and apply a correction for the projection contaminations using a Monte Carlo simulation method for SSA22-3.067, SSA22-3.093, PKS1138-2.160, BOSS1244-2.230, BOSS1244-2.246, and BOSS1542-2.241. 
This procedure has been successfully implemented to BOSS1244 and BOSS1542 \citep[see][for technical details]{Liu2023}, and we note that our pair selection criteria follow those in \cite{Liu2023}, with an additional criterion (3) applied to spectroscopically confirmed protocluster members.
Doing so, we subtract the $f_{\rm false\ pair}$ from $f_{\rm proj\ pair}$ and obtain the true pair fractions for each protocluster.

\begin{figure}
\includegraphics[width=\columnwidth]{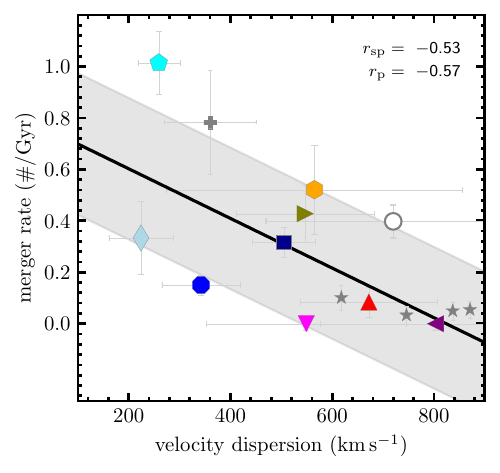}
\caption{Galaxy merger rate as a function of velocity dispersion for protoclusters at $z=2-3$. The symbols used are the same as in Fig.~\ref{fig2}. Also, the $r_{\rm sp}$ and $r_{\rm p}$ represent the correlation coefficients between galaxy merger rate and velocity dispersion (take PKS1138-2.160 as an exception).}
\label{fig3}
\end{figure}

We further estimate galaxy merger rate, from close pair fraction by dividing a specific time-scale. 
For massive galaxies with $\log(M_*/M_\odot) \geq 10.3$ and major companions within 5–30\,kpc, we adopt the merging time-scale from the {\textsc{Emerge}} simulation \citep{Oleary2021}, given by $T_{\rm obs} = -0.177 \times (1+z) + 1.205$. The calculated merger rates for each protocluster are listed in Table~\ref{tab1}.
We present the pair fraction and merger rate as a function of velocity dispersion for our sample protoclusters in Figs.~\ref{fig2} and \ref{fig3}. We apply the Spearman and Pearson tests and obtain correlation coefficients $r_{\rm sp}= -0.32$ ($r_{\rm sp}= -0.37$) and $r_{\rm p}=-0.48$ ($r_{\rm p}=-0.50$) between pair fraction (merger rate) and velocity dispersion. 
Here $r_{\rm p}$ assesses the linear relationship, while $r_{\rm sp}$ evaluates the monotonic relationship.
We emphasize that the identification of close pairs in PKS1138 is based on the rest-frame ultraviolet (UV) images (F814W), primarily contributed by young stellar populations with clumpy star formation regions. We find a stronger anti-correlation of $r_{\rm sp} = -0.53$ ($r_{\rm sp} = -0.53$) and $r_{\rm p} = -0.56$ ($r_{\rm p} = -0.57$) between pair fraction (merger rate) and velocity dispersion when excluding PKS1138-2.160.
The best-fitting linear functions are shown in Figs.~\ref{fig2} and \ref{fig3}, which indicates to a higher frequency of galaxy pairs (merger rates) in protoclusters with lower velocity dispersion, i.e. dynamically colder systems.

\subsection{Morphological Parameters}\label{sec3.4:morph}

\begin{figure}
\includegraphics[width=\columnwidth]{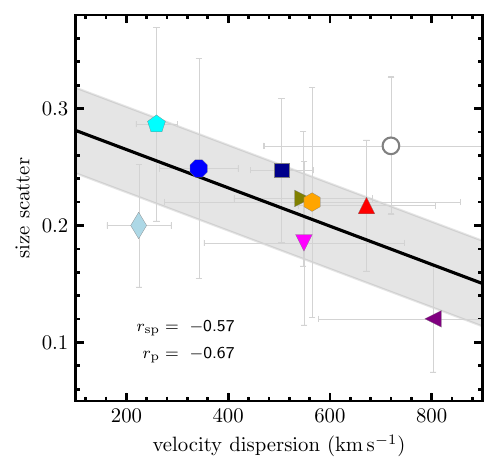}
\caption{The dispersion of galaxy half-light radii as a function of velocity dispersion for our sample of ten protoclusters. The symbols used are identical to those in  Fig.~\ref{fig2}. The parameters $r_{\rm sp}$ and $r_{\rm p}$ shows the correlation coefficients between size scatter and velocity dispersion, excluding PKS1138-2.160 (gray open circle).}
\label{fig4}
\end{figure}

The software tool \textsc{GALFIT} \citep{Peng2002, Peng2010} is utilized to fit galaxy images with two-dimensional S\'ersic models, and give the best-fitting half-light radius ($r_{\rm e}$) and S\'ersic index ($n$) through the least squares method. 
The majority of member galaxies in our selected protoclusters are star-forming galaxies and can be well fitted with S\'ersic models. 
We examine the stellar mass and size relation for the member galaxies of individual protoclusters and compare it with the stellar mass–size relation of star-forming galaxies in the general fields \citep{vanderWel2014}.
We fit a straight line to our data points in the $\log M_\ast - \log r_{\rm e}$ diagram at $\log M_\ast/{\rm M}_\odot \geq 10.3$ and quantify the scatter of the data points relative to the best-fitting relation. The size scatter, $\sigma_{\rm size}$, is calculated as the standard deviation in $r_{\rm e}$ relative to the best-fitting size-mass relation \citep{Newman2012, vanderWel2014}. It serves as a measure of the variation in the balance between dissipative and dissipationless formation processes integrated over cosmic history within protoclusters. 
We present the stellar mass-size relation for each protocluster, along with the histogram of size differences between protocluster members and the best-fitting relation, in Appendix~\ref{secB}.

Fig.~\ref{fig4} shows the size scatter as a function of velocity dispersion for our sample of ten protoclusters. 
It is worth noting that the half-light radii are measured from the rest-frame UV images (F814W) for the member galaxies of PKS1138-2.160, while the rest-frame optical images (F160W) are used for other protoclusters.
After excluding the PKS1138-2.160 data point, We show the best-fitting linear relation in Fig.~\ref{fig4}.  A clear anti-correlation is observed, with a Spearman correlation coefficient of $r_{\rm sp} = -0.57$ and a Pearson coefficient of $r_{\rm p} = -0.67$.
This strong anti-correlation indicates that the size scatter decreases with increasing velocity dispersion (i.e. the dynamical state of protoclusters becomes hotter). 

We also observe that $\sigma_{\rm size}$ derived from the rest-frame UV images for PKS1138-2.160 is larger than these from the rest-frame optical images for other protoclusters with similar dynamical states. 
The rest-UV light traces mostly young massive stars, while the rest-optical light is more dominated by older and less massive stars.  
These distinct stellar populations often have different spatial distributions, leading to color gradients within galaxies. These gradients are typically negative (red-to-blue, inside-out; \citealt{Suess2019a, Suess2019b}), though flat gradients also exist \citep{Jimenez-Andrade2021}. 
Moreover, the rest-frame UV light is heavily attenuated by dust, particularly in massive galaxies \citep{Martis2016, Cullen2017, Wilkins2017, Whitaker2017, Barger2023}. 
However, the dust attenuation relies not only on the dust surface density but also on the complex dust-star geometries \citep{Calzetti1994, Charlot2000, Li2019, Qin2024}.
These interactions could explain the larger $\sigma_{\rm size}$ based on the rest-UV images for PKS1138-2.160 at cosmic noon.

\section{Discussion and Summary} \label{sec4:discussion and summary}

In this study, we collect a pilot sample of ten protoclusters at $z=2$--3 with spectroscopic redshifts and HST imaging data to investigate the relationship between protocluster dynamical states and galaxy structures. 
We observe a higher merger rate in protoclusters with colder dynamical states, supported by correlation coefficients of  $r_{\rm sp} = -0.53$ and $r_{\rm p} = -0.57$. Additionally, we find another inverse correlation between size scatter and protocluster dynamics, with values of $r_{\rm sp} = -0.57$ and $r_{\rm p} = -0.67$. This suggests that frequent growth and compaction of galactic discs occur in protoclusters characterized by cold dynamical states.

\begin{figure}
\includegraphics[width=\columnwidth]{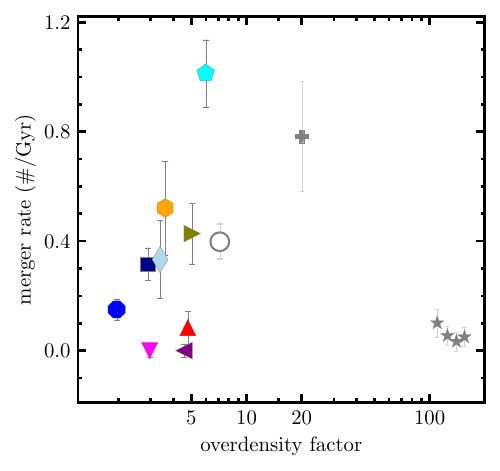}
\caption{Merger rates as a function of the overdensity factor for our protocluster sample. Symbols are the same as in Fig.~\ref{fig2}. The overdensity factors for galaxies are calculated within the contours shown in Fig.~\ref{fig1}. }
\label{fig5}
\end{figure}

\subsection{Increase of galaxy merger rate at increasing velocity dispersion}

Previous studies \citep[e.g.,][]{Lotz2013,Hine2016,Delahaye2017,Coogan2018, Watson2019,Liu2023} have presented controversial results on galaxy merger rates in protoclusters in comparison with that of the general field at the same epoch.  
Our measurements show the galaxy merger rate (i.e. true pair fraction) varies from one to another among our sample protoclusters. Importantly, the merger rate is anti-correlated with the velocity dispersion. In other words, protoclusters in a colder dynamical state host a higher frequency of galaxy mergers. Our finding explains the observed enhancement of galaxy merger rate in the protocluster IRC-0218A at $z = 1.62$, which is a dynamically-cold structure characterized by $\sigma_{\rm los} = 360\pm90$\kms \citep{Pierre2012, Lotz2013}.  Similarly,  four mature clusters at $1.59<z<1.71$ with $\sigma_{\rm los}>$ 500\kms were found to have a  pair fraction of $0.11^{+0.07}_{-0.06}$ \citep{Delahaye2017},  following the relation presented in Figs.~\ref{fig2} and \ref{fig3}.  
Furthermore, the pair fraction becomes lower at $\sigma_{\rm los} >$ 600\kms \ in the Abell clusters at $z<0.1$ from \cite{Sheen2012}, supporting our results of the anti-correlation between galaxy merger rate and velocity dispersion. 
In Figs.~\ref{fig2} and \ref{fig3}, we add  protocluster IRC-0218A and Abell clusters for comparison, and stress that the observed anti-correlation is applicable to (proto)clusters at $z<2$.

\cite{Shibuya2024} reported a linear relationship between merger rate and galaxy overdensity. This aligns with the expectation that higher galaxy densities may lead to more frequent mergers. 
We show the galaxy merger rate as a function of the overdensity factor for our sample of ten protoclusters and mature clusters in Fig.~\ref{fig5}, but do not find a linear trend.
Instead, our findings suggests that velocity dispersion plays a more significant role in determining merger rate, which could also explain the declined merger rate in mature clusters. In dynamically hot environments, galaxies are more likely to flyby rather than to merge, due to their relatively high encounter velocities. 
The enhancement of the merger rate in group-like environments with moderate overdensity factor and velocity dispersion is supported by the observational investigations \citep[e.g.,][]{Kocevski2011}.

Our finding of the decrease of galaxy merger rate in dynamically-hotter protoclusters can be used to interpret the enhancement of submillimetre galaxies in the outskirts of two protoclusters BOSS1244 and BOSS1542 \citep{Zhang2022}, and the emergence of quiescent galaxies in the dense core of BOSS1244 \citep[e.g.][]{Shi2024}. 
Galaxy mergers and interactions are considered as the physical driver triggering AGN activity in protoclusters \citep{Monson2023,Gatica2024} and the outskirts of relaxed clusters \citep{Koulouridis2024}, also playing a crucial role in the strong AGN feedback in PKS1138 \citep{Shimakawa2024}. 
Notably, the AGN fraction ($f_{\rm AGN}$) decreases at increasing $\sigma_{\rm los}$ in clusters of galaxies \citep{Popesso2006}. 
Our result on the anti-correlation between galaxy merger rate (or pair fraction) and $\sigma_{\rm los}$ implies that the triggering of AGNs is related to a higher rate of galaxy mergers   in dynamically-colder protoclusters.

\subsection{Increase of size scatter at decreasing velocity dispersion}

Multiple physical mechanisms  can drive the growth and shrinkage of galactic discs in gas-rich environments. 
The size of galaxy discs may increase through star formation in the cold gas accreted into the outer discs, while the discs may collapse (or shrink) due to disc instability  \citep{DB2014,Zolotov2015}.  At cosmic noon the gas accretion along filaments into protoclusters has been evidenced to be highly efficient \citep{Daddi2021}. The  complex large-scale tidal torques coupled with the surrounding structures leave their imprint on the inflowing gas along the filaments and  in turn regulate the angular momentum of galaxies and consequently their size \citep{Rost2021}.
Galaxy interactions and mergers induce rapid transformation in galaxy morphology and star formation enhancement/suppression, and thus play stronger roles in regulating the size growth of galaxies.  

A high frequency of these events may result in a large dispersion in galaxy size.  This dispersion is quantified by our parameter  `size scatter' ($\sigma_{\rm size}$)  for protocluster galaxies at given stellar masses. The anti-correlation between velocity dispersion and galaxy pair fraction naturally leads to a decrease of size scatter at increasing velocity dispersion, as shown in Fig.~\ref{fig4}.  In other words, a higher rate of galaxy mergers in dynamically-colder protoclusters is responsible for a larger size scatter. We examine the relationship between pair fraction and size scatter and
find a strong correlation characterized by  a Spearman coefficient $r_{\rm sp}=0.66$ and a Pearson  coefficient $r_{\rm p}=0.67$. Therefore, the two relationships of pair fraction and size scatter with velocity dispersion are consistent with each other.

In the literature, contradictory results can be found on the distribution of galaxy sizes in clusters when comparing to that of field galaxies \citep[e.g.][]{Kuchner2017,Afanasiev2023,Naufal2023}.  We attribute this contradiction largely to the variation in galaxy merger rate that is jointly regulated by the galaxy density (i.e. overdensity) and the dynamical state of protoclusters.  

Our results provide a first measurement of galaxy merger rate and size scatter across various dynamical states of protoclusters.  We emphasize that the dynamical state of protoclusters is a key physical quantity to understand the structural evolution of the member galaxies. Taking the effects of the dynamical states into account,  the  contradictory results about the properties of galaxies in protoclusters reported  in previous studies may be better understood in a consistent way.   Moreover, our findings indicate that the assembly stages of large-scale structures deeply impact the evolution of member galaxies. These results provide new insights into the evolutionary connections between the assembly of large-scale structures and galaxy formation.


\section*{Acknowledgements}
We thank the anonymous referee for her/his valuable comments and suggestions that improved this manuscript. 
This work is supported by the National Key Research and Development Program of China (2023YFA1608100),  the National Science Foundation of China (12233005, 12073078, 12273013, and 12303015), the science research grants from the China Manned Space Project with NO. CMS-CSST-2021-A02, CMS-CSST-2021-A04 and CMS-CSST-2021-A07, and the Chinese Academy of Sciences (CAS) through a China-Chile Joint Research Fund (CCJRF \#1809) administered by the CAS South America Center for Astronomy (CASSACA).  V.G. gratefully acknowledges support by the ANID BASAL project FB210003 and  from ANID FONDECYT Regular 1221310. DDS acknowledges financial support from the start-up funding of the Anhui University of Science and Technology (2024yjrc104) and the National Science Foundation of Jiangsu Province (BK20231106).

\section*{Data Availability}

The data underlying this article will be shared on reasonable request to the corresponding author.



\bibliographystyle{mnras}
\bibliography{ref}

\appendix

\section{stellar mass measurement for protocluster members in SSA22}\label{secA}

\cite{Kubo2015} estimated the stellar masses for 38 protocluster members by modeling their spectral energy distributions (SEDs) from the $u$-band to 8\,$\mu$m. 
However, only 7/15 galaxies in SSA22-3.067/SSA22-3.093 have stellar mass estimates using this method. 
We estimate stellar masses for HST/F160W-detected galaxies based on their luminosities and mass-to-light ratios \citep{Bell2003}. 
The rest-frame $u-g$ color at $z=3.09$ is derived from publicly available catalogues, including photometry in $R$ from \cite{Nestor2013}, $B$ and $V$ from \cite{Yamada2012}, $g$, $R$, and $K$ from \cite{Saez2015}, $u$, $B$, $V$, and $R_{\rm c}$ from \cite{Hayashino2019}, along with the HST/F160W data. We successfully derive stellar masses for 19/22 galaxies in SSA22-3.067/SSA22-3.093.

\begin{figure}
\includegraphics[width=0.95\columnwidth]{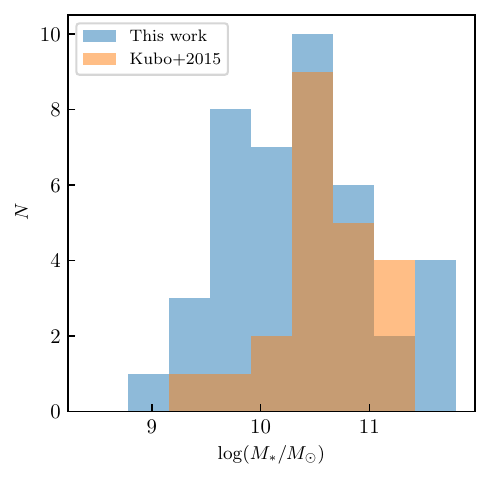}
\caption{The distribution of stellar masses for galaxies in protoclusters SSA22-3.067 and SSA22-3.093. Blue and orange represent the different stellar mass estimation methods from this work and \protect\cite{Kubo2015}, respectively.}
\label{figA1}
\end{figure}

The stellar mass distributions of protocluster members in SSA22-3.067 and SSA22-3.093 from both methods are shown in Fig.~\ref{figA1}. Notably, our analysis focuses on F160W-detected galaxies, whose stellar masses are primarily estimated using the mass-to-light ratio. Only two of these galaxies have SED-fitted stellar masses, differing by 0.22 dex from the mass-to-light ratio estimates. Therefore, we use the stellar masses derived from the mass-to-light ratio method in this paper and stress that using different methods does not affect our main conclusions.

\section{The Stellar Mass-Size Relation}\label{secB}

\begin{figure}
\begin{minipage}{0.498\textwidth}
\includegraphics[width=0.455\textwidth,trim=0 23 35 0, clip]{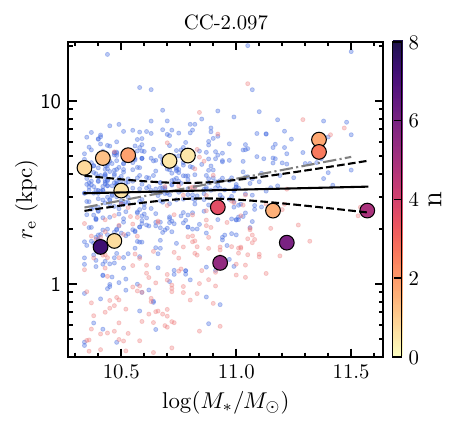}
\includegraphics[width=0.49\textwidth,trim=20 23 0 0, clip]{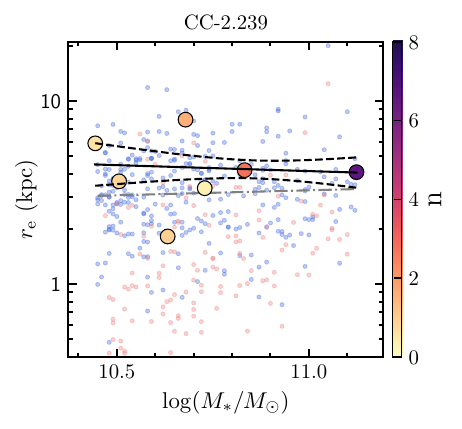}
\end{minipage}
\begin{minipage}{0.498\textwidth}
\includegraphics[width=0.455\textwidth,trim=0 23 35 0, clip]{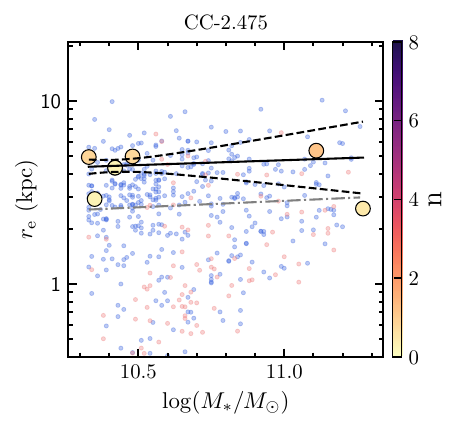}
\includegraphics[width=0.49\textwidth,trim=20 23 0 0, clip]{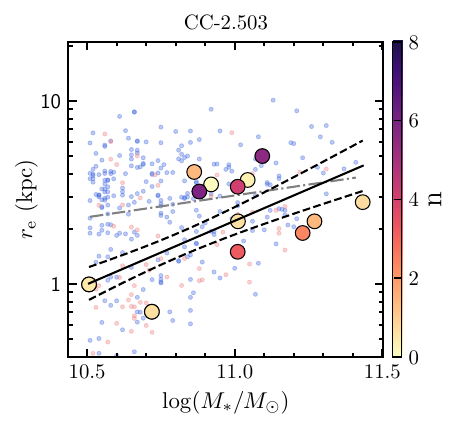}
\end{minipage}
\begin{minipage}{0.498\textwidth}
\includegraphics[width=0.455\textwidth,trim=0 23 35 0, clip]{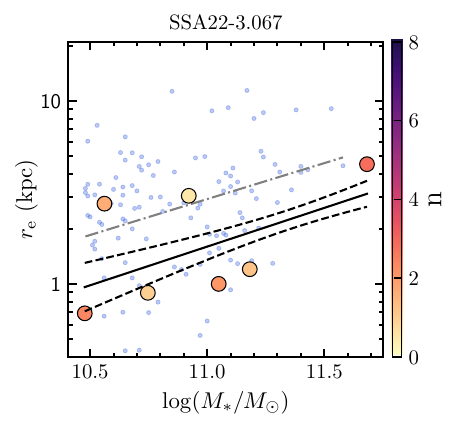}
\includegraphics[width=0.49\textwidth,trim=20 23 0 0, clip]{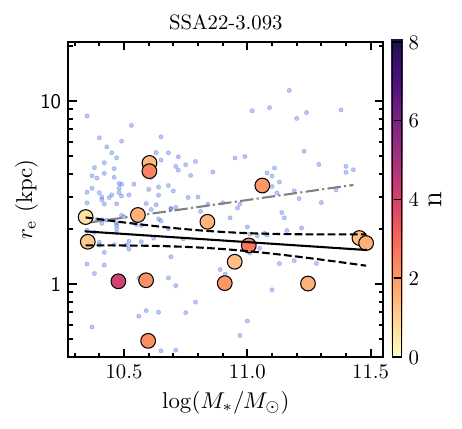}
\end{minipage}
\begin{minipage}{0.498\textwidth}
\includegraphics[width=0.455\textwidth,trim=0 23 35 0, clip]{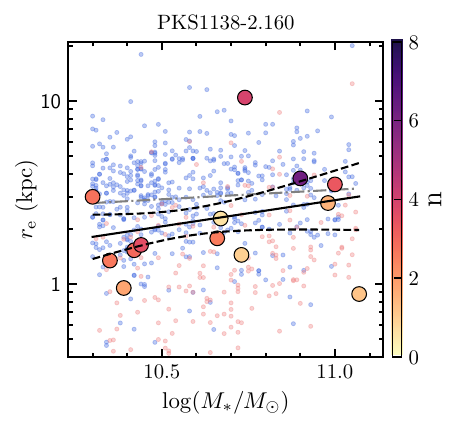}
\includegraphics[width=0.49\textwidth,trim=20 23 0 0, clip]{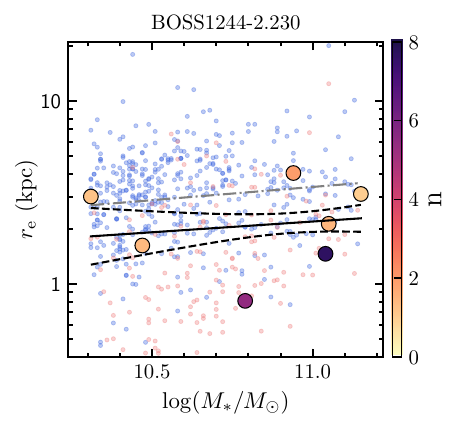}
\end{minipage}
\begin{minipage}{0.498\textwidth}
\includegraphics[width=0.455\textwidth,trim=0 0 35 0, clip]{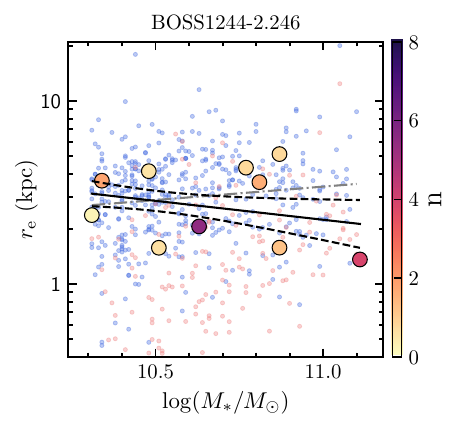}
\includegraphics[width=0.49\textwidth,trim=20 0 0 0, clip]{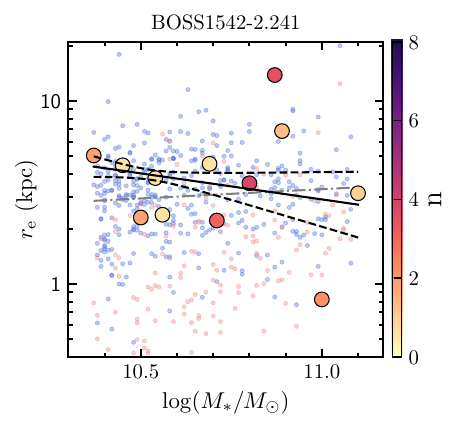}
\end{minipage}
    \caption{Distribution of protocluster members in the stellar mass versus half-light radius diagram is shown as filled circles, color-coded by S\'ersic indices. Blue and red dots represent field SFGs and QGs within $z_{\rm peak}\pm0.2$ of each protocluster. The gray dash-dotted line shows the best-fitting stellar mass–$r_{\rm e}$ relation for field SFGs, while the black solid and dashed lines represent the best-fitting relation and the corresponding standard deviation for protocluster members, resampled 100 times using the bootstrapping method.
    } \label{figB1}
\end{figure}

\begin{figure*}
\begin{minipage}{0.98\textwidth}
\includegraphics[width=0.208\textwidth,trim=0 0 0 0, clip]{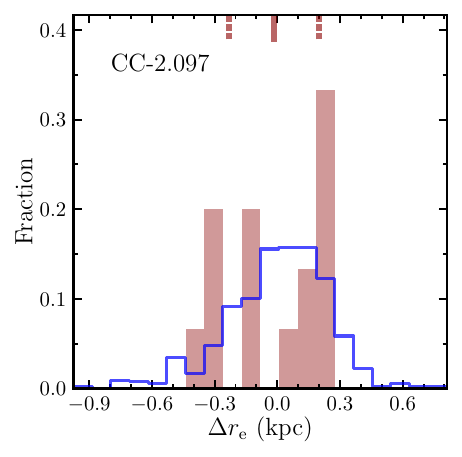}
\includegraphics[width=0.189\textwidth,trim=20 0 0 0, clip]{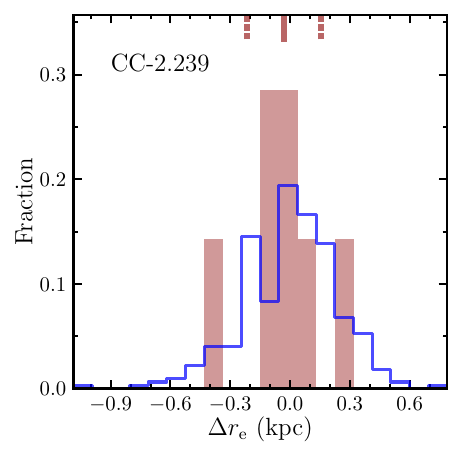}
\includegraphics[width=0.189\textwidth,trim=20 0 0 0, clip]{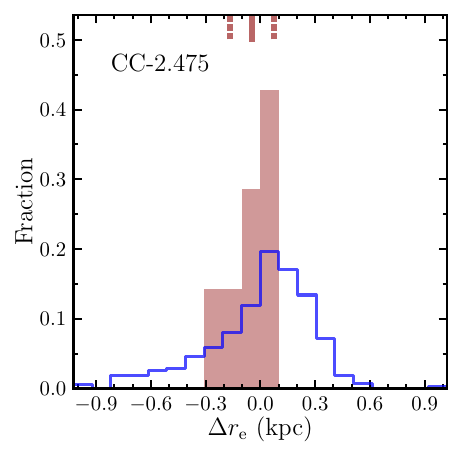}
\includegraphics[width=0.189\textwidth,trim=20 0 0 0, clip]{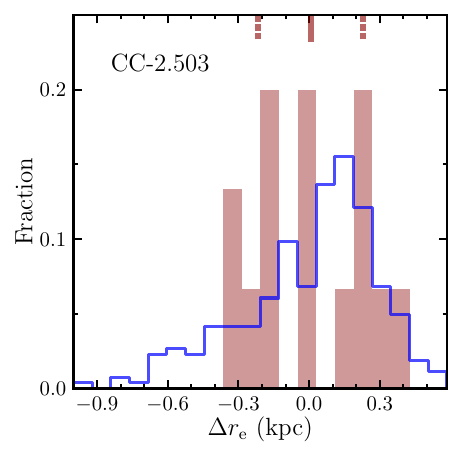}
\includegraphics[width=0.189\textwidth,trim=20 0 0 0, clip]{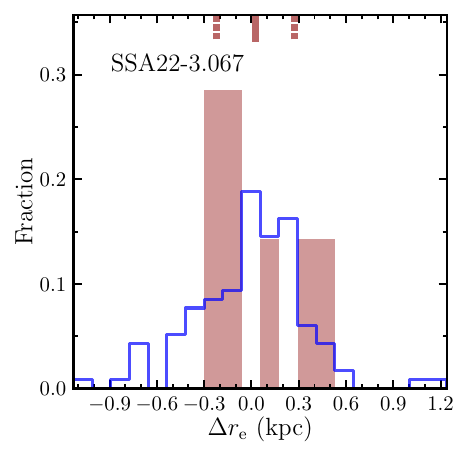}
\end{minipage}
\begin{minipage}{0.98\textwidth}
\includegraphics[width=0.208\textwidth,trim=0 0 0 0, clip]{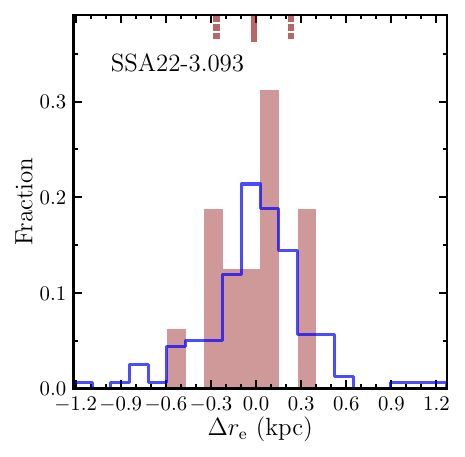}
\includegraphics[width=0.189\textwidth,trim=20 0 0 0, clip]{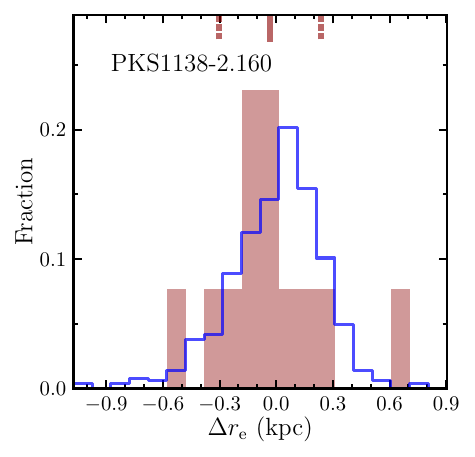}
\includegraphics[width=0.189\textwidth,trim=20 0 0 0, clip]{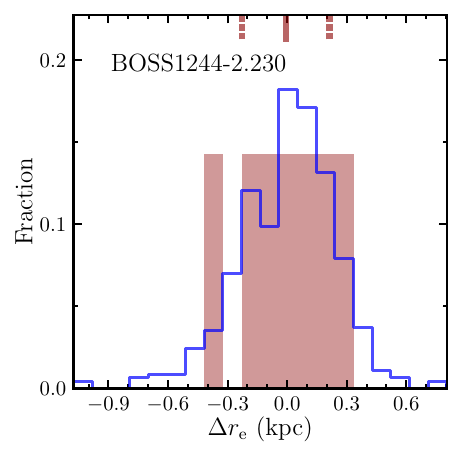}
\includegraphics[width=0.189\textwidth,trim=20 0 0 0, clip]{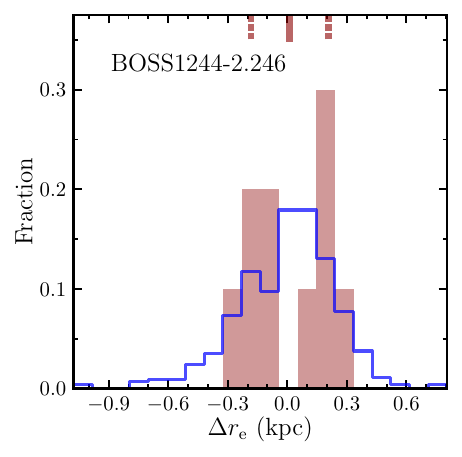}
\includegraphics[width=0.189\textwidth,trim=20 0 0 0, clip]{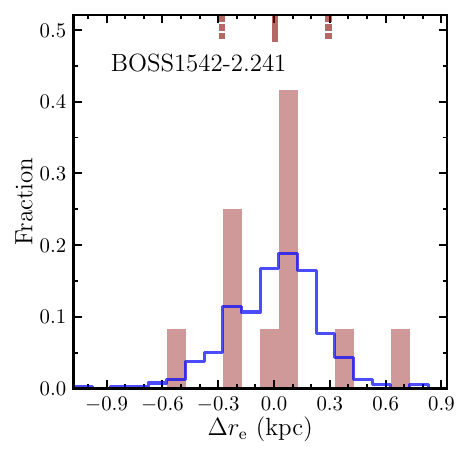}
\end{minipage}
    \caption{Distribution of size difference $\Delta r_{\rm e}$ for protocluster members and field SFGs, denoted by darkred and blue histograms. The vertical lines in the top of each panel show the mean value ($\Delta r_{\rm e, mean}$, solid line) and standard deviation (i.e., $\Delta r_{\rm e, mean} \pm \sigma_{\rm size}$, dotted lines) of protocluster members. } \label{figB2}
\end{figure*}

Here, we present the distribution of protocluster members in the stellar mass versus half-light radius ($r_{\rm e}$) diagram for our sample of ten protoclusters in Fig.~\ref{figB1}.
We construct a control sample of field star-forming and quiescent galaxies (SFGs and QGs) from the 3D-HST/CANDELS survey for each protocluster, matching the stellar mass range of protocluster members and constrained to $z_{\rm peak} \pm 0.02$ for each protocluster, shown as blue and red dots in Fig.~\ref{figB1}.
We use the bootstrapping method to resample the data 100 times, obtaining the linear function between the stellar mass and $r_{\rm e}$ in logarithm space for both protocluster member galaxies and field SFGs, as shown in each panel.

We then calculate the size difference, $\Delta r_{\rm e,i} = r_{\rm e,i}(M_,i) - r_{\rm e_{\rm best,i}}(M_,i)$, where $r_{\rm e,i}(M_,i)$ represents the half-light radius of each protocluster member, and $r_{\rm e_{\rm best,i}}(M_,i)$ is the best-fitting half-light radius at the corresponding stellar mass for each protocluster member.
We perform the same calculation for field SFGs to compare their size differences with those of the protocluster members.
We show their distributions with red and blue histograms in Fig.~\ref{figB2}. The mean value ($\Delta r_{\rm e, mean}$) and its standard deviation (i.e., $\Delta r_{\rm e, mean} \pm \sigma_{\rm size}$) is marked by short vertical lines in each panel.

\bsp	
\label{lastpage}
\end{document}